\documentclass[proof]{WileyASNA-v1}
\usepackage{lineno}
\articletype{Article Type}%

\begin{document}

\title{Further evidences of superluminal AGNs as $\gamma$-ray sources}

\author[1,2,3]{Hubing Xiao}
\author[1,4,6]{Junhui Fan*}
\author[2,3]{Riccardo Rando}
\author[1]{Jingtian Zhu}
\author[5]{Liangjun Hu}

\authormark{Hubing Xiao \textsc{et al}}

\address[1]{\orgdiv{Center for Astrophysics}, \orgname{Guangzhou University}, \orgaddress{\state{Guangzhou, 510006}, \country{China}}}
\address[2]{\orgdiv{Department of Physics and Astronomy "G. Galilei"}, \orgname{University of Padova}, \orgaddress{\state{Padova PD, 35131}, \country{Italy}}}
\address[3]{\orgdiv{INFN}, \orgname{University of Padova}, \orgaddress{\state{Padova PD, 35131}, \country{Italy}}}
\address[4]{\orgdiv{Astronomy Science and Technology Research Laboratory of Department of Education of Guangdong Province}, \orgname{Guangzhou University}, \orgaddress{\state{Guangzhou, 510006}, \country{China}}}
\address[5]{\orgdiv{School of physics and electronic information}, \orgname{Anhui normal University}, \orgaddress{\state{Wuhu, 241000}, \country{China}}}
\address[6]{\orgdiv{Key Laboratory for Astronomical Observation and Technology of Guangzhou}, \orgname{Guangzhou University}, \orgaddress{\state{Guangzhou, 510006}, \country{China}}}

\corres{*Junhui Fan, \email{fjh@gzhu.edu.cn}}

\abstract{In our previous work in \citet{Xiao2019}, we suggested that 6 superluminal sources could be $\gamma$-ray candidates, and in fact 5 of them have been confirmed in the fourth \textit{Fermi}-LAT source catalogue (4FGL). In this work, based on the 4FGL, we report a sample of 229 \textit{Fermi} detected superluminal sources (FDSs) including 40 new FDSs and 62 non-Fermi detected superluminal sources (non-FDSs). Thus, we believe that all superluminal sources should have $\gamma$-ray emissions, and superluminal motion could also be a clue to detect $\gamma$-ray emission from active galactic nuclei (AGN). We present a new approach of Doppler factor estimate through the study of the $\gamma$-ray luminosity ($L_{\gamma}$) and of the viewing angle ($\phi$).}


\maketitle


\section{introduction}\label{sec1}
Active galactic nuclei (AGNs) are a hot topic in astrophysics since they were discovered in the 1960s, while their nature is still object of investigation.
 Blazars, a very extreme subclass of AGNs,
 show rapid and high variability,
 high and variable polarization,
 variable and strong $\gamma$-ray emission and even superluminal motions
 (\citealt{vonMontigny1995};
  \citealt{Fan2013a};
  \citealt{Fan2013b};
  \citealt{Xiao2019}).
 These extreme observational properties of blazars are mainly due to the presence of a relativistic jet (\citealt{Blandford1979}).
  Blazars
  have two subclasses,
  namely BL Lacertae objects (BL Lacs) and
  flat spectrum radio quasars (FSRQs).
  Spectra of BL Lacs show weak or no emission lines while
  FSRQs show strong emission line features.

\textit{Fermi}-LAT (\textit{Fermi} Large Area Gamma-Ray Space Telescope), 
which was launched 2008, 
provides us with good opportunities to detect GeV $\gamma$-ray sources, for instance blazars. 
The third \textit{Fermi}-LAT source catalogue (3FGL), which draws up a list of the first 4 years data, contains 3034 $\gamma$-ray sources, 
out of  them 1459 are blazars.
The \textit{Fermi} detected blazar sample is enlarged by the fourth \textit{Fermi}-LAT source catalogue (4FGL), 
 which concludes with 2938 blazars out of 5098 $\gamma$-ray sources from 8 years data (\citealt{Fermi2019}). 
The $\gamma$-ray detections of blazars can be used to investigate the emission mechanism of the high energetic $\gamma$-ray emissions. 
The correlations between $\gamma$-ray luminosity and other band luminosities 
 (radio, optical, and X-ray emissions) were discussed by many authors in literature 
 (\citealt{Dondi1995}; 
 \citealt{Fossati1998}; 
 \citealt{Fan2016}; 
 \citealt{ZF2018}). 
It was found that the correlation between $\gamma$-ray and radio emission is the result of 
a self-synchrotron Compton (SSC) emission. 
The evidence that the $\gamma$-ray emission in blazars is strong also suggests the existence of a relativistic beaming effect
 (\citealt{Arshakian2010};
  \citealt{Fan2013a};
  \citealt{Fan2013b};
  \citealt{Fan2017};
  \citealt{Giovannini2014};
  \citealt{Giroletti2012};
  \citealt{Kovalev2009};
  \citealt{Massaro2013a};
  \citealt{Massaro2013b};
  \citealt{Savolainen2010};
  \citealt{Xiao2015};
  \citealt{Pei2016};
  \citealt{Yang2017};
  \citealt{Yang2018a};
  \citealt{ZF2018};
  \citealt{Xiao2019}).

An interesting observational phenomenon in blazars is the presence of superluminal motions (sources with apparent velocity larger than the speed of light, $\beta_{\rm app} > 1$), which is also a consequence of the beaming effect.
In 1996, \citeauthor{Fan1996} compiled a sample of 48 superluminal sources in order to investigate the beaming effect and found that the core dominance parameter, the ratio of core radio flux to the extended radio flux, is an indicator of the orientation of the emission.
They proposed that the superluminal motion and beaming effect are probably two aspects of the same phenomenon.
\citet{ZF2008} collected a sample of 123 superluminal sources and found that the radio emissions are strongly boosted by the beaming effect and that the superluminal motion is an another indicator of the beaming effect in AGNs.
\citet{Lister2009} found that there is an overwhelming tendency to display outward motions, studying the radio flux of 135 radio-loud AGNs extracted from the MOJAVE \footnote{http://www.physics.purdue.edu/MOJAVE/} (Monitoring of Jets in Active galactic nuclei with VLBA Experiments) sample.
\citet{Lister2013} studied jet orientation variations and superluminal motions of 200 AGNs and found a general trend of increasing apparent speed with increasing distance along the jet for both radio galaxies and BL Lac objects.
In \citet{Xiao2019}, we collected 291 superluminal sources, including 189 FDSs (\textit{Fermi} Detected $\gamma$-ray Superluminal sources) and 102 non-FDSs (non-\textit{Fermi} Detected Superluminal sources) to study the difference between FDSs and non-FDSs. 
We found that FDSs are more strongly beamed than non-FDSs, and we claimed that some superluminal sources are $\gamma$-ray emission candidates. 
This suggestion is further confirmed by the data of the 4FGL release (\citealt{Fermi2019}), which allow to extend our previous study. 

This work is arranged as follows:
In section 1, we introduce the background of $\gamma$-ray blazars and of the superluminal motion;
in section 2, we give our sample and results;
discussion and conclusions are presented in sections 3 and  4.

\section{Samples and Results}\label{sec2}
\subsection{Samples}
In \citet{Xiao2019}, we identified, in the 3FGL catalogue, 291 superluminal sources, out of them which 189 FDSs and 102 non-FDSs.

We matched the sources in our non-FDSs sample with the sources in the 4FGL catalogue and found that 40 of 102 sources that were non-FDSs in 3FGL are FDSs in 4FGL.
They are listed in Table \ref{Tab-new-FDS-simp}. 
Hence, we have an updated sample of 229 FDSs from 3FGL and 4FGL, and remain 62 non-FDSs.

Meanwhile, the corresponding source information, optical magnitude, radio flux, X-ray flux density, 3FGL $\gamma$-ray flux density, from \citet{Xiao2019} will be also employed in this work. 
In Table \ref{Tab-new-FDS-simp}, we only listed the 40 new FDSs,
column (1) the source 4FGL name; 
column (2) other name; 
column (3) classification, $F$ = FSRQs, $B$ = BL Lacs, $Sy$ = Seyfert galaxy, $G$ = galaxy and $U$ = BCU, (blazar candidates of uncertain type) through this paper;  
column (4) redshift; 
column (5) and (6) radio Doppler factor and its corresponding reference, H09 is \citet{Hovatta2009}, L18 means \citet{Liodakis2018}; 
column (7), (8) and (9) are the integral photon flux and uncertainty from 1 to 100 GeV and its photon index from 4FGL.

\begin{center}
\begin{table*}[t]%
\centering
\caption{New FDSs from 4FGL.\label{Tab-new-FDS-simp}}%
\tabcolsep=0pt%
\begin{tabular*}{36pc}{@{\extracolsep\fill}lcccccccc@{\extracolsep\fill}}
\toprule
4FGL name 		& Other name 	 & Class   & redshift  		&    $\delta_R$ & Ref.       &    Flux1000                    &  Flux1000\_error          & $\alpha_{\rm ph}$    \\
		   		& 		         & 	        &               	&                      	&              &    $photons/cm^{2}/s$    &  $photons/cm^{2}/s$   & 		                        \\
      (1)	  		&     (2)	         &    (3)     &    (4)       	&    (5)              &   (6)       &    (7)                               &     (8)   		              &	(9)			       \\
\midrule	
4FGL J0006.3-0620	&	0003-066	&	B		&	0.347	&	6.96		&	L18	&	1.26E-10	&	3.35E-11	&	2.1704	\\
4FGL J0013.6+4051	&	0010+405	&	F		&	0.255	&	8.81		&	L18	&	1.81E-10	&	3.52E-11	&	2.212	\\
4FGL J0019.6+7327	&	0016+731	&	F		&	1.781	&	7.84		&	L18	&	1.27E-09	&	7.91E-11	&	2.5943	\\
4FGL J0037.6+3653	&	0035+367	&	F		&	0.366	&			&		&	1.67E-10	&	3.63E-11	&	2.3478	\\
4FGL J0109.7+6133	&	0106+612 &	G		&	0.783	&	23.74	&	L18	&	2.56E-09	&	1.15E-10	&	2.6068	\\
4FGL J0112.0+3442	&	0109+351	&	F		&	0.45		&	1.57		&	L18	&	1.51E-10	&	3.34E-11	&	2.3388	\\
4FGL J0115.1-0129	&	0112-017 	&	F		&	1.365	&	14.55	&	L18	&	2.44E-10	&	4.02E-11	&	2.7209	\\
4FGL J0125.7-0015	&	0122-003 	&	F		&	1.074	&	20		&	L18	&	1.22E-10	&	3.01E-11	&	3.0257	\\
4FGL J0152.2+2206	&	0149+218	&	F		&	1.32		&	4.32		&	L18	&	3.75E-10	&	4.65E-11	&	2.7059	\\
4FGL J0210.7-5101	&	0208?512	&	F		&	1.003	&			&		&	4.19E-09	&	1.09E-10	&	2.3536	\\
4FGL J0228.7+6718	&	0224+671	&	F		&	0.53		&	6.01		&	L18	&	5.05E-10	&	6.64E-11	&	2.545	\\
4FGL J0231.8+1322	&	0229+131 &	F		&	2.059	&	11.31	&	L18	&	4.48E-10	&	5.26E-11	&	2.7412	\\
4FGL J0359.6+5057	&	0355+508 &	F		&	1.52		&	6.12		&	L18	&	2.29E-09	&	1.29E-10	&	2.6416	\\
4FGL J0403.3+2601	&	0400+258 &	F		&	2.109	&	7.59		&	L18	&	2.32E-10	&	4.63E-11	&	2.6		\\
4FGL J0433.0+0522	&	0430+052	&	F		&	0.033	&	1.09		&	L18	&	6.94E-10	&	7.56E-11	&	2.7163	\\
4FGL J0539.6+1432	&	0536+145 &	F		&	2.69		&	22.59	&	L18	&	8.14E-10	&	7.78E-11	&	2.5573	\\
4FGL J0555.6+3947	&	0552+398	&	F		&	2.365	&	25.2		&	H09	&	6.63E-10	&	7.13E-11	&	2.8018	\\
4FGL J0728.0+6735	&	0723+679	&	Sy1.2	&	0.846	&			&		&	1.08E-10	&	2.60E-11	&	3.0044	\\
4FGL J0746.0-0039	&	0743-006 	&	F		&	0.994	&	4.7		&	L18	&	1.42E-10	&	3.91E-11	&	2.629	\\
4FGL J0748.6+2400	&	0745+241	&	F		&	0.409	&	1.81		&	L18	&	6.74E-10	&	5.59E-11	&	2.2817	\\
4FGL J0808.5+4950	&	0804+499	&	F		&	1.435	&	8.63		&	L18	&	2.26E-10	&	3.38E-11	&	2.8056	\\
4FGL J0836.5-2026	&	0834-201	&	F		&	2.752	&			&		&	1.81E-10	&	4.88E-11	&	2.8311	\\
4FGL J0850.0+5108	&	0846+513 &	F		&	0.585	&	22.77	&	L18	&	2.16E-09	&	7.90E-11	&	2.2701	\\
4FGL J0910.0+4257	&	0906+430	&	F		&	0.67		&			&		&	2.41E-10	&	3.61E-11	&	2.5156	\\
4FGL J0958.0+4728	&	0955+476	&	F		&	1.882	&	12.01	&	L18	&	3.87E-10	&	4.22E-11	&	2.6414	\\
4FGL J1037.4-2933	&	1034-293 	&	F		&	0.312	&	2.8		&	F09	&	3.58E-10	&	4.79E-11	&	2.4824	\\
4FGL J1051.6+2109	&	1049+215 &	F		&	1.3		&	6.29		&	L18	&	1.68E-10	&	3.51E-11	&	2.7773	\\
4FGL J1131.0+3815	&	1128+385 &	F		&	1.74		&			&		&	7.83E-10	&	6.03E-11	&	2.5512	\\
4FGL J1131.4-0504	&	1128-047 	&	G		&	0.266	&	0.51		&	L18	&	3.11E-10	&	5.72E-11	&	2.4798	\\
4FGL J1159.3-2142	&	1157-215 	&	F		&	0.927	&			&		&	1.17E-09	&	7.37E-11	&	2.5131	\\
4FGL J1459.0+7140	&	1458+718	&	Sy 1.5	&	0.905	&	3.92		&	L18	&	2.14E-10	&	2.84E-11	&	2.4462	\\
4FGL J1534.8+0131	&	1532+016	&	F		&	1.42		&	10.13	&	L18	&	1.03E-09	&	6.79E-11	&	2.3663	\\
4FGL J1638.1+5721	&	1637+574	&	F		&	0.75		&	7.97		&	L18	&	2.98E-10	&	3.25E-11	&	2.6426	\\
4FGL J1753.7+2847	&	1751+288	&	F		&	1.115	&	3.81		&	L18	&	1.68E-10	&	3.70E-11	&	2.3828	\\
4FGL J2011.6-1546	&	2008-159	&	F		&	1.18		&	7.75		&	L18	&	1.96E-10	&	4.84E-11	&	2.841	\\
4FGL J2038.7+5117	&	2037+511	&	F		&	1.686	&	19.86	&	L18	&	1.28E-09	&	1.21E-10	&	2.616	\\
4FGL J2123.6+0535	&	2121+053	&	F		&	1.941	&	10.31	&	L18	&	4.33E-10	&	5.15E-11	&	2.3652	\\
4FGL J2219.2-0342	&	2216-038	&	F		&	0.901	&	12.23	&	L18	&	1.74E-10	&	4.04E-11	&	2.8928	\\
4FGL J2225.6+2120	&	2223+210 &	F		&	1.959	&	11.26	&	L18	&	2.05E-10	&	3.81E-11	&	2.8114	\\
4FGL J2354.6+4554	&	2351+456	&	F		&	1.992	&	3.36		&	L18	&	2.22E-10	&	4.11E-11	&	2.4002	\\
\bottomrule
\end{tabular*}
\end{table*}
\end{center}

\subsection{Results: Correlation between $\gamma$-ray band luminosity and other band luminosities for FDSs}
We studied the correlations between $\gamma$-ray luminosity and luminosity in other wavelength ranges (radio, optical, X-ray) for the FDSs. 
The multi-wavelength (radio, optical and X-ray) data are collected from the BZCAT, the $\gamma$-ray photon flux ($photons/s/cm^{2}$) are collected from 3FGL and 4FGL. 
The optical magnitude has been corrected by Galactic extinction correction and transformed to flux density. Then, the radio, optical, X-ray and $\gamma$-ray flux densities has been all multiply by $(1+z)^{\alpha-1}$ in order to introduce the $K$-correction.
For the spectral indices ($\alpha$),
 we adopt $\alpha_{\rm r}=0$ in the radio band (\citealt{Donato2001}, \citealt{Abdo2010}), and in the X-ray band
 $\alpha_{\rm X}=0.78$ for FSRQs,
 $\alpha_{\rm X}=1.30$ for BLs and
 $\alpha_{\rm X}=1.05$ for BCUs (blazar candidates of uncertain type) as suggested by \citet{Fan2016}. 

Considering that the luminosity is correlated with the redshift, we have removed its influence in order to achieve the pure correlation of the luminosities through a partial correlation analysis:
$${\rm r}_{ij,k} = \frac{{\rm r}_{ij} -{\rm r}_{ik}{\rm r}_{jk}}{\sqrt{(1 - {\rm r}_{ik}^2)(1 - {\rm r}_{jk}^2)}},$$ 
here ${\rm r}_{ij}, \  {\rm r}_{ik}, \ {\rm r}_{jk}$ are the correlation coefficients between any pair of variables $i$, $j$ and $k$. In this work, the variables $i$ and $j$ represent luminosities of two bands, the variable $k$ represents redshift. 

Any study of blazar radiation must take into account the presence of Doppler beaming effect, in order to identify correlations between luminosities in different spectral bands. The correlation between observed (beamed) flux, $f^{\rm ob}$, and intrinsic (de-beamed) flux, $f^{\rm in}$, is $f^{\rm in} = f^{\rm ob} / \delta^{p}$, where, $\delta$ is the Doppler factor, and $p = 2 + \alpha$ for continuous flow, $p = 3 + \alpha$ for discrete flow, respectively. The correlation between observed (beamed) luminosity and intrinsic (de-beamed) luminosity is $L^{\rm in} = L^{\rm ob} / \delta^{1+p}$. 
The optical, $\delta_{\rm o}$, and the X-ray, $\delta_{\rm X}$, Doppler factors, were derived from the radio Doppler factor, $\delta_{\rm r}$, through an empirical method proposed by \citet{Fan1993}:
$\delta_{\rm \nu}=\delta_{\rm o}^{1+1/8~ {\rm log} ({\nu}_{\rm o}/{\nu})}$,
where $\delta_{\rm o}$ is the optical Doppler factor,
and $\delta_{\rm X} \sim \delta_{\rm o}^{0.5}$,
$\delta_{\rm r} \sim \delta_{\rm o}^{1.5}$ are corresponding factors of the X-ray and radio ranges respectively.
The $\gamma$-ray Doppler factor is the same as the radio Doppler factor, because the radio and $\gamma$-ray emission arise from the same region in a leptonic model.

\begin{figure*}[t]
\centerline{\includegraphics[width=5.5in]{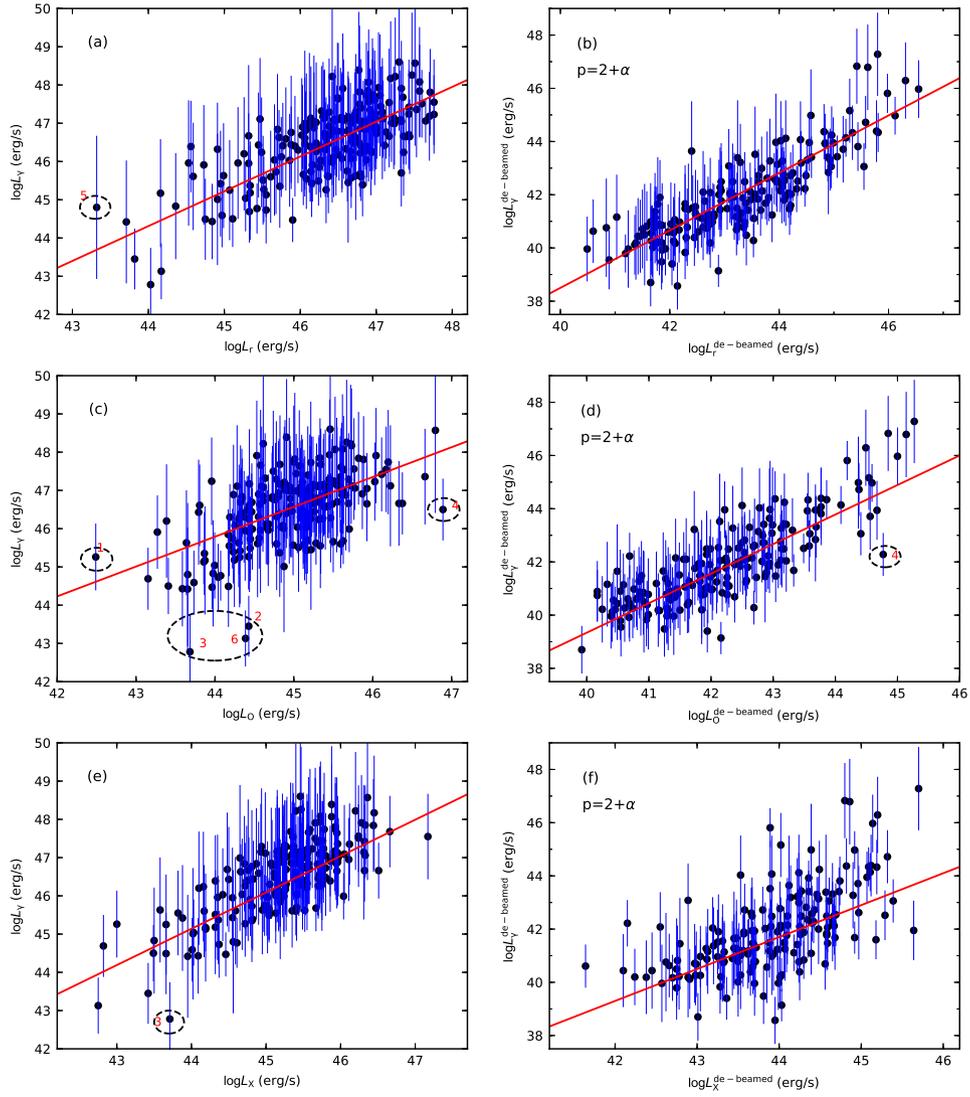}}
\caption{Plots for $\gamma$-ray band and other bands luminosity correlations for FDS. Left-hand panel for beamed results, right-hand panel for de-beamed results. Upper panel for $\gamma$-ray band against radio band, middle panel for $\gamma$-ray against optical and lower panel for $\gamma$-ray against X-ray band. Black dots with blue error-bar, calculated from the corresponding uncertainties that collected from 3FGL and 4FGL, for the FDS sources, and the red solid lines stand for best linear fitting results.\label{L_gamma-L}}
\end{figure*}

Figure \ref{L_gamma-L} reports the distribution of pairs of observed beamed luminosities (left column) and the distribution of pairs of corresponding de-beamed luminosities (right column). The dots in the encircled regions have been excluded from the evaluation of the best fit of the distributions on the base of considerations reported in section 3.2.
Table \ref{G-F} quotes the following parameters: $a$ and $b$ are slope and intercept of the linear best fit; $\Delta a$ and $\Delta b$ are the corresponding uncertainties; $N$ is the total number of sources, the number in parentheses represents the number of sources we neglected in the regression; $r$ and $p$ are the correlation coefficient and the chance probability, respectively.    

\begin{center}
\begin{table*}[t]
\caption{$\gamma$-ray flux luminosity $vs$ other bands luminosities partial correlation analysis results.\label{G-F}}
\centering
\begin{tabular*}{35pc}{@{\extracolsep\fill}cccccccc@{\extracolsep\fill}}
\hline
 \ Type \ & \ band \ & \ Type \ & \ $a+ \Delta a$ \ & \ $b+\Delta b$ \ & \ N \ & \ r \ & \ p \ \\   \hline
 & ${\rm log}L_{\rm \gamma} ~vs~ {\rm log}L_{\rm r}$ & beamed & $0.91 \pm 0.06$ & $4.27 \pm 2.65$ & 213+(1) &  0.14  &  4.0\%   \\   \cline{3-8}
 & & de-beamed & $1.07 \pm 0.05$ & $-4.70 \pm 2.09$ & 180 &  0.87  &  $2.2 \times 10^{-55}$   \\   \cline{2-8}
 FDS & ${\rm log}L_{\rm \gamma} ~vs~ {\rm log}L_{\rm o}$ & beamed & $0.78 \pm 0.07$ & $11.47 \pm 3.18$ &  209+(5)  &  0.25  &  $2.2 \times 10^{-4}$    \\   \cline{3-8}
 & & de-beamed & $1.11 \pm 0.06$ & $-5.06 \pm 2.60$ & 178+(1) &  0.82  &  $1.8 \times 10^{-45}$   \\   \cline{2-8}
 & ${\rm log}L_{\rm \gamma} ~vs~ {\rm log}L_{\rm X}$ & beamed & $0.95 \pm 0.06$ & $3.34 \pm 2.57$ & 201+(1) &  0.33  &  $2.3 \times 10^{-6}$    \\   \cline{3-8}
 & & de-beamed & $1.20 \pm 0.13$ & $-11.10 \pm 5.68$ & 170 &  0.75   &  $1.3 \times 10^{-31}$   \\   \hline
\end{tabular*}
\end{table*}
\end{center}

\section{Discussion}
\subsection{New sources}
In our previous work in \citet{Xiao2019}, we predicted 6 non-FDS sources to be $\gamma$-ray candidates, 0153+744, 0208-512, 0536+145, 0552+398, 2223+210, 2351+456, with a criterion that $\beta^{\rm max}_{\rm non-FDS} > \langle \beta^{\rm max}_{\rm FDS}\rangle + 5\sigma$ (\citealt{Xiao2019}). Out of these 6 candidates, 5 sources have been identified through 4FGL as 4FGL J0210.7-5101 (0208-512), 4FGL J0539.6+1432 (0536+145), 4FGL J0555.6+3947 (0552+398), 4FGL J2225.6+2120 (2223+210) and 4FGL J2354.6+4554 (2351+456). 

In addition, we found 40 non-FDS sources from \citet{Xiao2019} as new FDS sources in 4FGL. Thus, for 291 available superluminal sources, 229 are Fermi-detected sources, namely 78.7 \% of the superluminal sources, while the remaining 62 (21.3\%) have no available Fermi detection. We suspect that all these non-FDS sources quite likely have a faint $\gamma$-ray emission, which lays under the level of the detectability of \textit{Fermi}-LAT.

\subsection{Correlations}
Correlations between the $\gamma$-ray luminosity and their corresponding luminosity in the radio, optical and X-ray bands for FDS sources are shown in Figure \ref{L_gamma-L}. Their numerical features are listed in Table \ref{G-F}. 
The left panel, Figure \ref{L_gamma-L}a, c, e, shows the correlations of ${\rm log}L_{\rm \gamma} \ vs \ {\rm log}L_{\rm r, \ o, \ X}$, where the luminosities are boosted by Doppler beaming effect. Meanwhile, the right panel, Figure \ref{L_gamma-L}b, d, f, shows the correlations of the de-beamed luminosities.

Six sources, 
(1) 3FGL J0205.0+1510 (4C +15.05),
(2) 3FGL J0325.2+3410,
(3) 4FGL J0433.0+0522,
(4) 3FGL J0710.5+4732 (S4 0707+47),
(5) 3FGL J1104.4+3812 (Mrk 421), and 
(6) 3FGL J1517.6-2422 (AP Lib)
were removed before calculating the linear regression.
In the subplot (a), only one sources, Mrk 421 was removed. This source is a typical high synchrotron peak frequency source (HSP) and shows relatively lower radio band luminosity.
In subplot (c), 5 sources are circled. Source 1 has low optical luminosity. Source 4 has both high observed and high intrinsic luminosity in subplots (c) and (d). Sources 2, 3, 6 have relatively low $L_\gamma$, among them sources 2 and 3 are both Seyfert galaxies, and source 6 is a BL Lac object. 
In subplot (e), only source 3 is neglected.
We did not circle any points in subplot (f), since the scatters are relatively well distributed separately in this subplot.

In Figure \ref{L_gamma-L}a, c, e, where the beaming effect is still present the correlation coefficients: 0.14, 0.25 and 0.33 indicate weak correlation of luminosities. The same coefficients become 0.87, 0.82, 0.75 respectively after corrections of beaming effect, namely dividing the luminosities by $\delta^{1+p}$, see Figure \ref{L_gamma-L}b, d, f.   

Therefore, the beamed observed luminosities show weak correlation, whereas the de-beamed intrinsic luminosities show strong correlation. A result which suggests that the radio, optical and X-ray emissions are all beamed.

\subsection{$\gamma$-ray Luminosity and Viewing Angle}
Considering the presence of a strong beaming effect in these superluminal sources, especially FDSs, we expected some good correlation between $\gamma$-ray Luminosity ($L_{\rm \gamma}$) and Viewing Angle ($\phi$). Therefore, we collected 181 FDSs with available Doppler factor from literature ($\delta_{\rm Ref}$) and maximum apparent velocity of all components, which are the individual bright features apart from the core region, from \citet{Xiao2019}. 
These data are listed in Table \ref{phi-L_gamma_gamma}, where we give: 
column (1) name;
column (2) other name;
column (3) classification;
column (4) redshift;
column (5) Doppler factor, $\delta_{\rm Ref}$; 
column (6) reference for $\delta_{\rm Ref}$;
column (7) 3FGL integral photon flux from 1 to 100 GeV; 
column (8) $\gamma$-ray photon spectral index; 
column (9) maximum apparent velocity;
column (10) estimated Doppler factor of this work, $\delta^{\rm L_\gamma}_{\rm TW}$. 

\begin{center}
\begin{table*}[t]%
\centering
\caption{181 FDSs with available Doppler factor.\label{phi-L_gamma_gamma}}%
\tabcolsep=0pt%
\begin{tabular*}{40pc}{@{\extracolsep\fill}cccccccccc@{\extracolsep\fill}}
\toprule
Fermi name & Other name  & Class  & redshift  & $\delta_{\rm Ref}$ & Ref. & Flux1000 & $\alpha_{\rm ph}$  & $\beta_{\rm app}^{\rm max}$ & $\delta^{\rm L_{\gamma}}_{\rm TW}$ \\
       &        &         &        &       &       & $photons/cm^{2}/s$ &       & $c$    & 		 \\
(1)   & (2)  & (3)    & (4)  & (5)  & (6) & (7)                             & (8) & (9)      & (10)	 \\
\midrule
3FGL J0006.4+3825	&	 S4 0003+38 	&	F	&	0.229	&	5.23		&	L18	&	6.06E-10	&	2.62	&	4.6	&	1.33 		\\
3FGL J0051.0-0649	&	0048-071 		&	F	&	1.975	&	5.61		&	L18	&	1.40E-09	&	2.10	&	13.4	&	37.32	\\
3FGL J0102.8+5825	&	TXS 0059+581	&	F	&	0.644	&	18.51	&	L18	&	6.18E-09	&	2.09	&	8.62	&	18.97	\\
3FGL J0108.7+0134	&	 4C+01.02 	&	F	&	2.099	&	2.64		&	L18	&	5.27E-09	&	2.26	&	25.8	&	69.03	\\
3FGL J0112.8+3207	&	0110+318 	&	F	&	0.603	&	10.21	&	L18	&	3.04E-09	&	2.36	&	19	&	9.91		\\
...				&	... 			&	...	&	...		&	... 		&	...	&	...		&	...	&	...	&	...		\\
\bottomrule
\end{tabular*}
\end{table*}
\end{center}

\begin{figure}[htbp]
 \centering
 \includegraphics[width=3.5in]{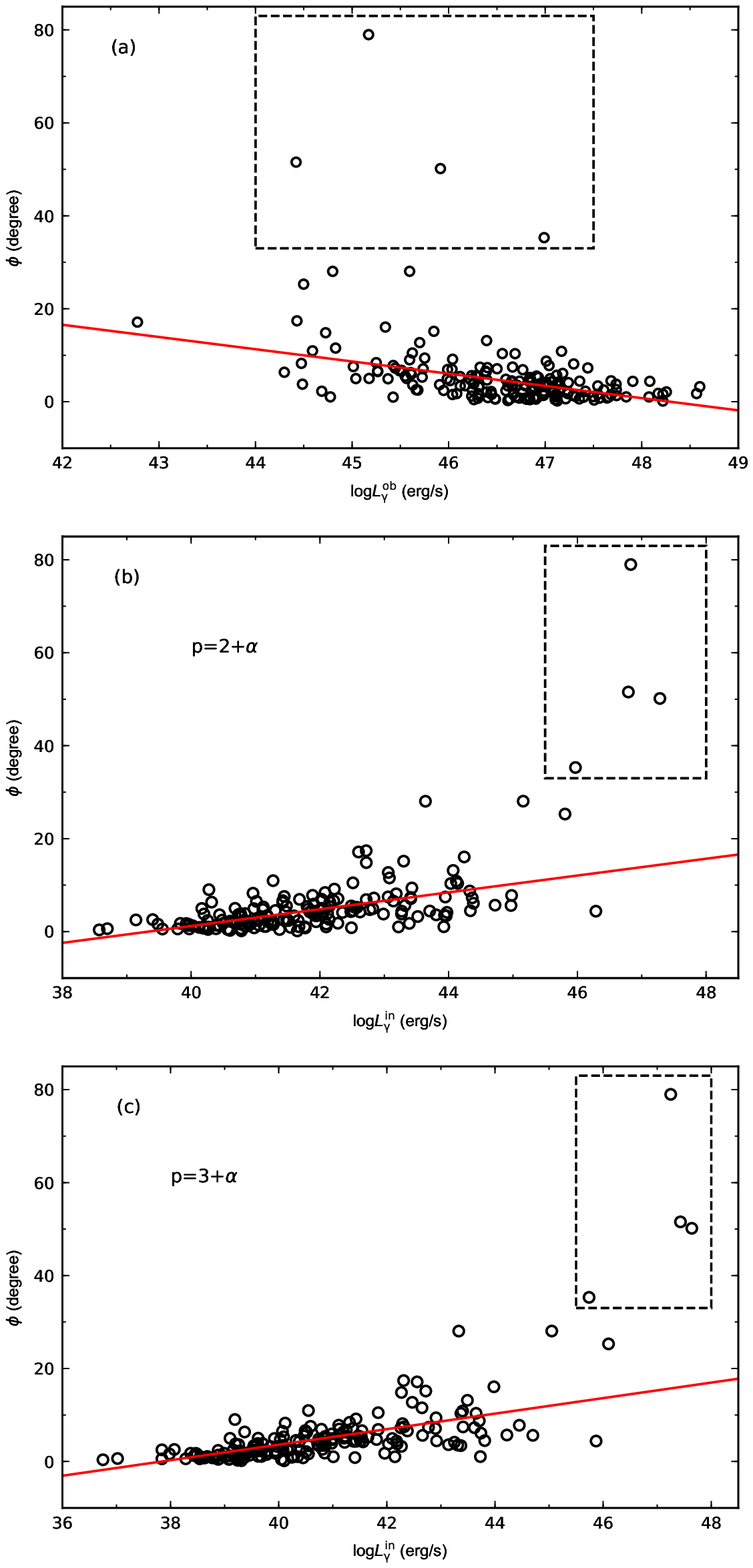}
  \caption{Correlation between observed $\gamma$-ray luminosity and viewing angle of FDS sources. The solid line shows the best linear fitting result of all the sources (apart from the sources in the dashed rectangle).}
  \label{phi-L_gamma}
\end{figure}

We found a significant anti-correlation, see Figure \ref{phi-L_gamma}a, between observed $\gamma$-ray luminosity ($L_{\rm \gamma}$) and viewing angle ($\phi$),
$$\phi=-(2.63 \pm 0.31) {\rm log}L_{\rm \gamma}^{\rm ob}+(127.02 \pm 14.48),$$ with $r=-0.54~{\rm and}~p= 1.1 \times 10^{-14}$. 

While, there is a positive correlation between intrinsic $\gamma$-ray luminosity ($L_{\rm \gamma}^{\rm in}$, which can be determined with $L_{\rm \gamma}^{\rm in} = L_{\rm \gamma}^{\rm ob} / \delta^{1+p}$) and viewing angle ($\phi$),
$$\phi=(1.81 \pm 0.20) {\rm log}L_{\rm \gamma}^{\rm in}+(-71.20 \pm 8.28),~(p=2+\alpha)$$ with $r=0.57~{\rm and}~p= 1.3 \times 10^{-16}$. 
$$\phi=(1.67 \pm 0.15) {\rm log}L_{\rm \gamma}^{\rm in}+(-63.19 \pm 6.16),~(p=3+\alpha)$$ with $r=0.64~{\rm and}~p= 8.8 \times 10^{-22}$. 
We show the last two correlations in Figure \ref{phi-L_gamma}b, c: the circles indicate the sources, and the solid red line represents the best linear regression. Four sources (3FGL J1015.0+4925, 3FGL J1058.6+5627, 3FGL J1510.9-0542, 3FGL J1653.9+3945), which lie within a rectangle in Figure \ref{phi-L_gamma} a, b, c have been neglected in the derivation of the linear regressions.

The opposite correlations of ${\rm log}L_{\rm \gamma}$ and $\phi$ indicate a strong beaming effect. If we hold these two correlations, then an estimate of the Doppler factor can be obtained as of function of $\gamma$-ray observed luminosity and the photon index. We use the two correlations as formulas couple with $L_{\rm \gamma}^{\rm in} = L_{\rm \gamma}^{\rm ob} / \delta^{1+p}$ and $\alpha = \alpha_{\rm ph}-1$, to derive a formula, which allows to estimate the Doppler factor:
$$\delta^{\rm L_\gamma} = 10^{\frac{2.45logL_{\rm \gamma}^{\rm ob}-109.51}{3+(\alpha_{\rm ph}-1)}}, ~(p=2+\alpha), ~ {\rm and} $$
$$\delta^{\rm L_\gamma} = 10^{\frac{2.57logL_{\rm \gamma}^{\rm ob}-113.90}{4+(\alpha_{\rm ph}-1)}}, ~(p=3+\alpha) $$

\begin{figure}[t]
\centerline{\includegraphics[width=3.5in]{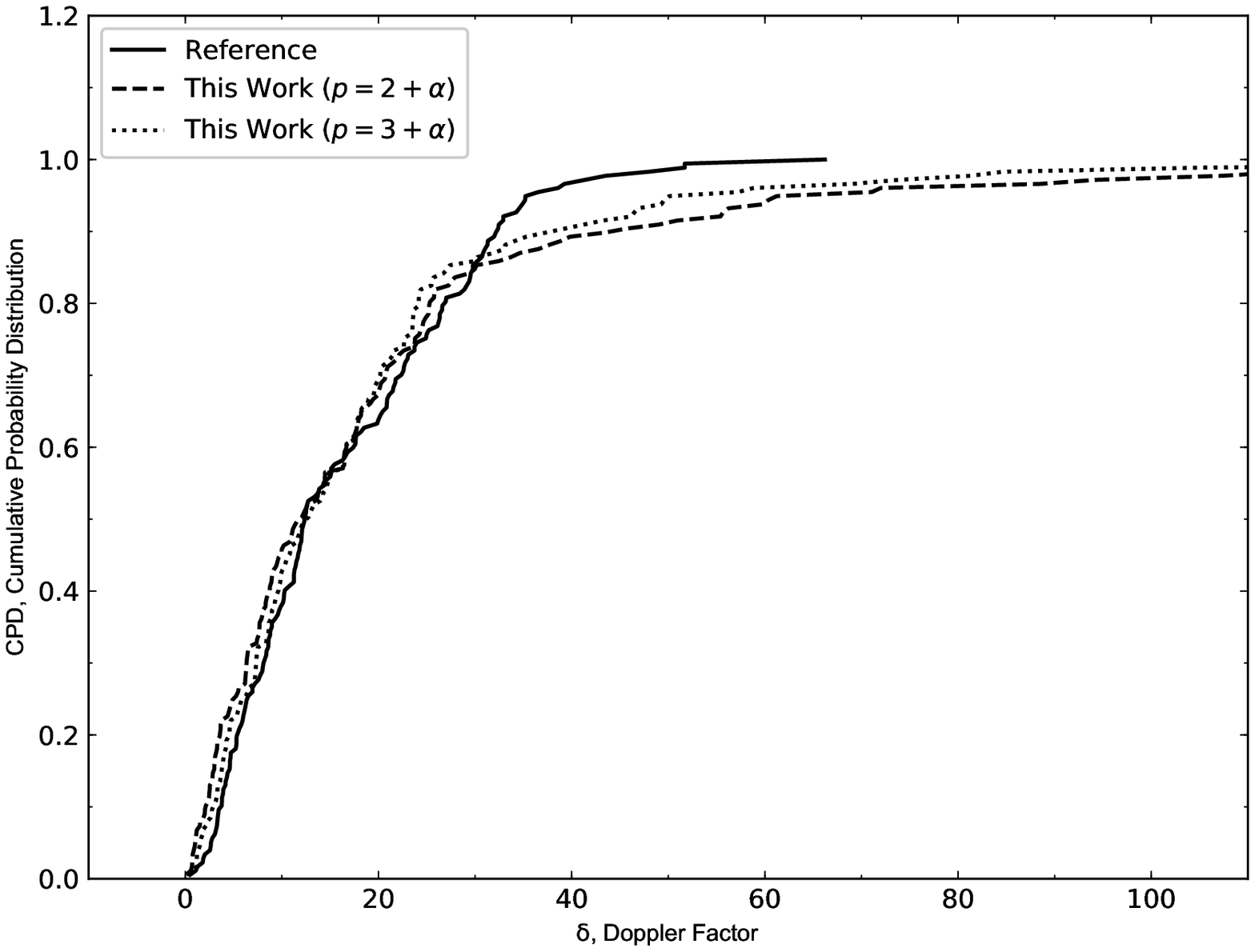}}
\caption{ K-S test CPD distribution. The solid line for the Doppler factor from literature, while dashed line and dotted line for Doppler factors are calculated in this work with different expression of $p$.\label{delta_ref-delta_TW34}}
\end{figure}

With this formula, we calculated the Doppler factor for the 181 FDSs, and listed the estimated Doppler factor $\delta^{\rm L_\gamma}_{\rm TW}$ in the last column in Table \ref{phi-L_gamma_gamma}. Then the K-S test between $\delta_{\rm Ref}$ and $\delta^{\rm L_\gamma}_{\rm TW}$ for each case, apart from the 4 sources in the rectangle, was applied to distinguish the two distributions. The results are listed in Table \ref{ks}, which quotes:
column (1) the condition applied to choose sources from the sample of 181 FDSs;
column (2) $p = 2 + \alpha$ and $p = 3 + \alpha$ for the continuous flow and  the discrete flow respectively;
column (3) classification;
column (4) number of sources;
columns (5) and (6) report the statistics value and the p-value from the K-S test results.
The results corresponding to a $statistics=0.07$ and $p-value=71.0\%$ for the case $p = 3+\alpha$, in Table \ref{ks} and Figure \ref{delta_ref-delta_TW34}, show that it is reasonable to affirm that the distributions of the two samples are the same.

\begin{center}
\begin{table*}[t]%
\centering
\caption{K-S test of $\delta_{\rm Ref}$ vs $\delta^{\rm L_\gamma}_{\rm TW}$.\label{ks}}%
\tabcolsep=0pt%
\begin{tabular*}{30pc}{@{\extracolsep\fill}cccccc@{\extracolsep\fill}}
\toprule
Filter &  p    &  Class  &  N   &  statistics & p-value     \\
(1)     & (2)  &   (3)      & (4)  &  (5)          & (6)  		\\
\midrule
						&				&	 All 		&	177		&	0.12		&	15.4 \%     \\
 						&	$2+\alpha$	&	 FSRQ 	&	142		&	0.11		&	31.0 \%     \\
with $\phi < 30^{\rm \circ}$	&				&	 BL Lac 	&	29		&	0.24		&	32.1 \%     \\  \cline{2-6}
(deduct 4 sources)			&				&	 All 		&	177		&	0.07		&	71.0 \%     \\
						&	$3+\alpha$	&	 FSRQ 	&	142		&	0.09		&	57.1 \%     \\
						&				&	 BL Lac 	&	29		&	0.21		&	51.4 \%     \\ 					
\bottomrule
\end{tabular*}
\end{table*}
\end{center}

By using our `luminosity-Doppler factor' formula, we calculated $\delta^{\rm L_\gamma}$ for 4FGL blazars, with available redshift from literature, and listed the results in Table \ref{1505}, where we quote:
column (1) name;
column (2) classification;
column (3) redshift;
column (4) 4FGL Integral photon flux from 1 to 100 GeV;
column (5) 4FGL $\gamma$-ray photon spectral index;
column (6) estimated luminosity-Doppler factor from this work.
This $\delta^{\rm L_\gamma}$ is neither a upper limit nor a lower limit of a real Doppler factor, but we may use it as comparison for further Doppler factor calculations and beaming effect studies. 

\begin{center}
\begin{table*}[t]%
\centering
\caption{1505 blazars luminosity-Doppler factor.\label{1505}}%
\tabcolsep=0pt%
\begin{tabular*}{30pc}{@{\extracolsep\fill}cccccc@{\extracolsep\fill}}
\toprule
Fermi name & Class  & Redshift  & Flux1000                   & $\alpha_{\rm ph}$     &    $\delta^{\rm L_\gamma}$      	\\
                    &            &                & $photons/cm^{2}/s$   &                                  &    							\\
       (1)         &   (2)    &      (3)      &      (4)                         &      (5)                        &       (6)					         \\
\midrule
4FGLJ0001.5+2113	&	F	&	1.106	&	9.6832E-10	&	2.6802	&	10.66	\\
4FGLJ0003.9-1149	&	B	&	1.30999	&	2.8624E-10	&	2.1267	&	9.78		\\
4FGLJ0004.0+0840	&	U	&	2.057129	&	2.2296E-10	&	2.1032	&	15.62	\\
4FGLJ0004.4-4737	&	F	&	0.88		&	4.8554E-10	&	2.4152	&	6.46		\\
4FGLJ0005.9+3824	&	F	&	0.229	&	4.5453E-10	&	2.6661	&	1.16		\\
...				&	...	&	...		&	...			&	...		&	...		\\
\bottomrule
\end{tabular*}
\end{table*}
\end{center}

Meanwhile, we applied both beaming model with $p = 2 + \alpha$ and $p = 3 + \alpha$ valid in the case of continuous flow and discrete flow respectively and found, comparing $\delta_{\rm ref}$ with $\delta^{\rm L_\gamma}_{\rm TW}$, that the emission of FDSs is most likely dominated by the discrete flow as indicated by the K-S test result in Table \ref{ks}.
This result suggests that the $\gamma$-ray emission from jets occurs not continuously, but acts separately in $\gamma$-ray emission, like when a relativistic flow of matter interacts through a shock with interstellar medium, accelerates electrons to relativistic speeds and these electrons scatter soft photons to $\gamma$-ray energy in knotty regions, called also `hot spots' or `knots'.

In order to investigate the origin of $\gamma$-ray photons, we collected the average total flux density at 15 GHz from \citet{Lister2018} and the component flux density at 15 GHz from \citet{Lister2019} for 187 sources, and listed them in Table \ref{187}, in which we quote:
column (1) name;
column (2) classification;
column (3) redshift;
column (4) the serial number of an individual component; 
column (5) the 15 GHz component flux density of the corresponding component; 
column (6) the 15 GHz total flux density of the source; 
column (7) $\gamma$-ray photon spectral index;
column (8) integral photon flux from 1 to 100 GeV;

\begin{center}
\begin{table*}[t]%
\centering
\caption{187 FDSs with total and component flux from \citet{Lister2018} and \citet{Lister2019}.\label{187}}%
\tabcolsep=0pt%
\begin{tabular*}{40pc}{@{\extracolsep\fill}cccccccc@{\extracolsep\fill}}
\toprule
Fermi name & Class  & Redshift  & Component ID & $f_{\rm 15 GHz}^{\rm com}$  & $f_{\rm 15 GHz}^{\rm tot}$ & $\alpha_{\rm ph}$ & Flux1000  		  \\
		   &		 &		   &  			     &    $mJy$  				     & 		$mJy$			 &  				& $photons/cm^{2}/s$ \\
(1)		   & (2)       & (3)           & (4) 		     & (5)					     & (6)					 & (7) 			& (8)				   \\
\midrule
3FGL J0051.0-0649	&	F	&	1.975	&	1	&	28	&	1115.4	&	2.1047	&	1.40E-09	\\
3FGL J0102.8+5825	&	F	&	0.644	&	5	&	154	&	2530.31	&	2.0943	&	6.18E-09	\\
3FGL J0108.7+0134	&	F	&	2.107	&	2	&	92	&	2303.69	&	2.2601	&	5.27E-09	\\
3FGL J0110.2+6806	&	B	&	0.29		&	1	&	32	&	328.8	&	1.9909	&	1.92E-09	\\
3FGL J0112.8+3207	&	F	&	0.603	&	3	&	21	&	640.44	&	2.3626	&	3.04E-09	\\
...				&	...	&	...		&	...			&	...		&	...		&	...		\\
\bottomrule
\end{tabular*}
\end{table*}
\end{center}

\begin{figure}[t]
\centerline{\includegraphics[width=3.7in]{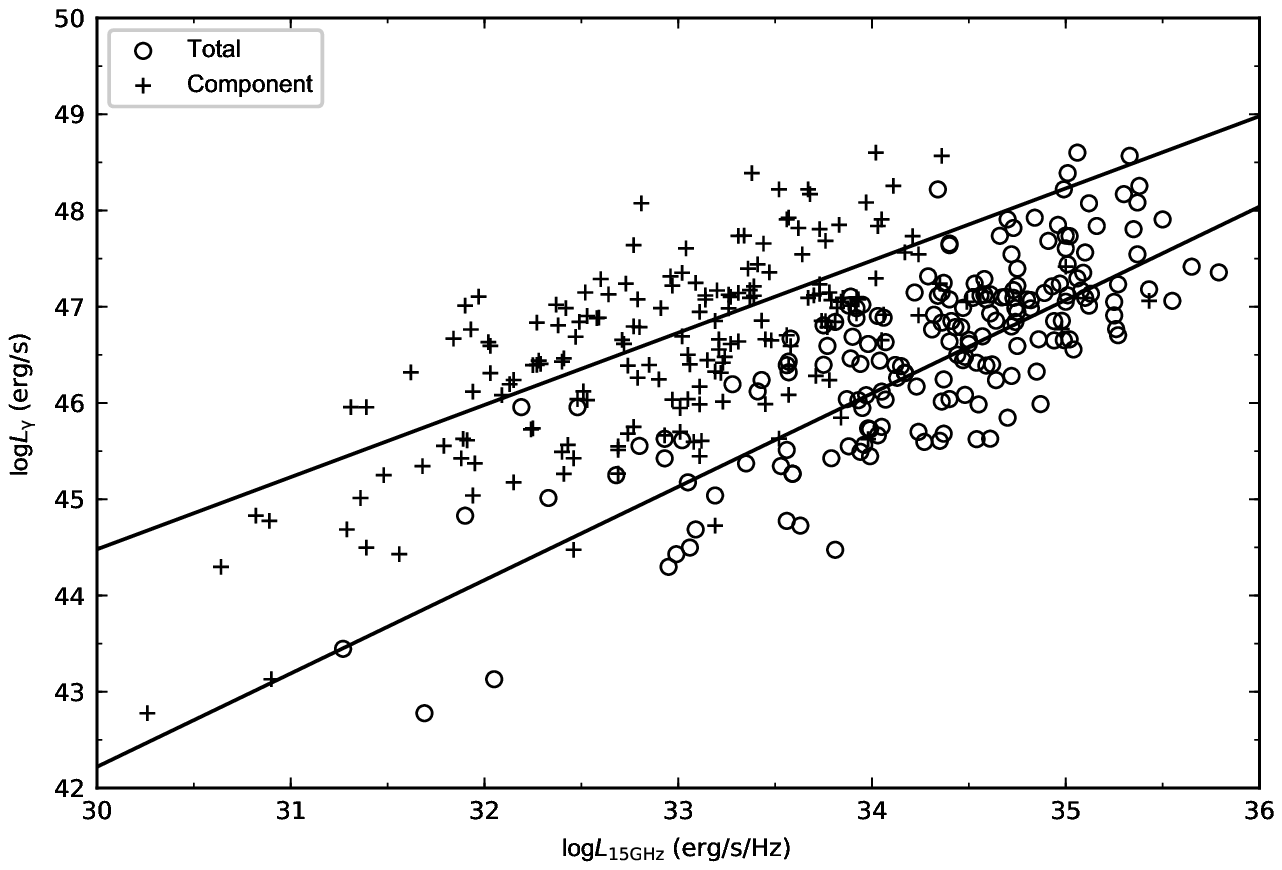}}
\caption{Correlation between $\gamma$-ray luminosity and radio 15 GHz luminosity, the open dots and the crosses represent the total and the component luminosities respectively.\label{L_com-L_tot-L_gamma}}
\end{figure}

The correlations between $L_{\rm \gamma}$ against 15 GHz total luminosity ($L_{\rm 15 GHz}^{\rm tot}$) and component luminosity ($L_{\rm 15 GHz}^{\rm com}$) are shown in Figure \ref{L_com-L_tot-L_gamma} and give the following parameterized relation,
$${\rm log} L_{\gamma} = (0.75 \pm 0.06) {\rm log} L_{\rm 15 GHz}^{\rm com} + (21.98 \pm 1.93)$$ with $ r = 0.68 ~{\rm and}~p = 3.6 \times 10^{-27};$
$${\rm log} L_{\gamma} = (0.97 \pm 0.06) {\rm log} L_{\rm 15 GHz}^{\rm tot} + (13.12 \pm 1.96)$$ with $ r = 0.78 ~{\rm and}~p = 6.5 \times 10^{-40}.$

These correlations are both good according to their $r$ and $p$ values. The correlation of $L_{\rm \gamma}$ against $L_{\rm 15 GHz}^{\rm com}$ indicates that a considerable amount of $\gamma$-ray photons originate from `hot spots' or `knots', where particles, accelerated to relativistic speeds, scatter soft photons towards high energies.

\section{Conclusions}
1. We have updated 40 new FDS sources from our 102 non-FDS sources through cross-matching our sample with the 4FGL catalogue;

2. We have found that the non-Fermi detected superluminal sources are $\gamma$-ray emitting candidates, this result helps to enlarge the $\gamma$-ray AGNs sample;

3. We have proposed that a Doppler factor estimate can be obtained through an empirical formula valid for $\gamma$-ray sources, especially for FDSs;

4. We suggest that the $\gamma$-ray emission from superluminal sources is discrete and that significant fraction $\gamma$-ray emission should arise from `knots' and `hot spots'.

\section*{Acknowledgements}
We would like to thank the nice referee, Prof. Rafanelli, for his corrections and notations made in the manuscript, and his comments and suggestions that let us improve the manuscript. This work was carried out under the auspices of the \fundingAgency{National Natural Science Foundation of China} under Contract No. \fundingNumber{NSFC 11733001 and NSFC U1531245}, and partially supported by \fundingAgency{Natural Science Foundation of Guangdong Province} under Contracts No. \fundingNumber{2019B030302001} and \fundingNumber{2017A030313011}, support for Astrophysics Key Subjects of Guangdong Province and Guangzhou City, and \fundingAgency{Science and Technology Program of Guangzhou} under Contract No. \fundingNumber{201707010401}. We also thank the MOJAVE team and Purdue University for the use of the precious kinematic data.


\begin{thebibliography}

\bibitem[Abdo et al.(2010)] {Abdo2010} Abdo, A. A., Ackermann, M., Agudo, I., et al. 2010, ApJ, 716, 30.

\bibitem[Arshakian et al.(2010)] {Arshakian2010} Arshakian, T. G., Torrealba, J., Chavushyan, V. H., et al. 2010, A\&A, 520, A62.

\bibitem[Blandford \& K\"onigl. et al.(1979)] {Blandford1979} Blandford, R. D., K\"onigl, A. 1979, ApJ, 1, 232.

\bibitem[Donato et al.(2001)] {Donato2001} Donato, D., Ghisellini, D., Tagliaferri, G and Fossati, G. 2001, A\&A, 375, 739.

\bibitem[Dondi \& Ghisellini(1995)] {Dondi1995} Dondi, L \& Ghisellini, G. 1995, MNRAS, 273, 583.

\bibitem[Fan et al.(1993)] {Fan1993} Fan, J. H., Xie, G. Z., Li, J. J. 1993, ApJ, 415, 113.

\bibitem[Fan et al.(1996)] {Fan1996} Fan, J. H., Xie, G. Z., Wen, S. L. 1996, A\&AS, 116, 409.

\bibitem[Fan et al.(2013a)] {Fan2013a} Fan, J. H., Yang, J. H., Liu, Y., Zhang, J. Y. 2013a, RAA, 13, 259.

\bibitem[Fan et al.(2013b)] {Fan2013b} Fan, J. H., Yang, J. H., Zhang, J. Y. et al. 2013b, PASJ, 65, 25.

\bibitem[Fan et al.(2016)] {Fan2016} Fan, J. H., Yang, J. H., Liu, Y.,  et al. 2016, ApJS, 226, 20.

\bibitem[Fan et al.(2017)] {Fan2017} Fan, J. H., Yang, J. H., Xiao, H. B., Lin, C., et al. 2017, ApJL, 835, 38.

\bibitem[{\textit{Fermi}} LAT collaboration.(2019)] {Fermi2019} {\textit{Fermi}} LAT collaboration. 2019, arXiv:1902.10045v2.

\bibitem[Fossati et al.(1998)] {Fossati1998} Fossati, G., Maraschi, L., Celotti, A., et al. 1998, MNRAS, 299, 433.

\bibitem[Giovannini et al.(2014)] {Giovannini2014} Giovannini, G., Liuzzo, E., Boccardi, B., Giroletti, M. 2014, IAUS, 304, 200.

\bibitem[Giroletti et al.(2012)] {Giroletti2012} Giroletti, M., Pavlidou, V., Reimer, A., et al. 2012, AdSpR, 49, 1320.

\bibitem[Hovatta et al.(2009)] {Hovatta2009} Hovatta, T., Valtaoja, E., Tornikorski, M., et al. 2009, A\&A, 496, 527.

\bibitem[Kovalev et al.(2009)] {Kovalev2009} Kovalev, Y. Y., et al. 2009, ApJ, 707, 56.

\bibitem[Liodakis et al.(2018)] {Liodakis2018} Liodakis, I., Hovatta, T., Huppenkothen, D., Kiehlmann, S., et al. 2018, ApJ, 886, 2.

\bibitem[Lister et al.(2009)] {Lister2009} Lister, M. L., Cohen, M. H., Homan, D. C., et al. 2009, AJ, 138, 1874.

\bibitem[Lister et al.(2013)] {Lister2013} Lister, M. L., Aller, M. F., Aller, H. D., Homan, D. C., et al. 2013, AJ, 146, 120.

\bibitem[Lister et al.(2018)] {Lister2018} Lister, M. L., Aller, M. F., Aller, H. D., et al. 2018, ApJ, 234, 12.

\bibitem[Lister et al.(2019)] {Lister2019} Lister, M. L., Homan, D. C., Hovatta, T., et al. 2019, ApJ, 874, 1.

\bibitem[Massaro et al.(2013a)] {Massaro2013a} Massaro, F., Giroletti, M., Paggi, A., et al. 2013a, ApJS, 207, 4.

\bibitem[Massaro et al.(2013b)] {Massaro2013b} Massaro, F., D'Abrusco, R., Giroletti, M., et al. 2013b, ApJS, 208, 15.

\bibitem[Pei et al.(2016)] {Pei2016} Pei, Z. Y., Fan, J. H., et al. 2016, Ap\&SS, 361, 237.

\bibitem[Savolainen et al.(2010)] {Savolainen2010} Savolainen, T., Homan, D. C., Hovatta, T., et al. 2010, A\&A, 512A, 24.

\bibitem[von Montigny et al.(1995)] {vonMontigny1995} von Montigny, C., Bertsch, D. L., et al. 1995, ApJ, 440, 525.

\bibitem[Xiao et al.(2015)] {Xiao2015} Xiao, H. B., Fan, J. H., Pei, Z. Y., et al. 2015, Ap\&SS, 359, 39.

\bibitem[Xiao et al.(2019)] {Xiao2019} Xiao, H. B., Fan, J. H., et al. 2019, SCPMA, 62, 129811.

\bibitem[Yang et al.(2017)] {Yang2017} Yang, J. H., Fan, J. H., Liu, Y., Zhang, Y. L., Yang, R. S., Tuo, M. X., et al. 2017, Ap\&SS, 362, 219.

\bibitem[Yang et al.(2018a)] {Yang2018a} Yang, J. H., Fan, J. H., Liu, Y., Zhang, Y. L., Yang, R. S., Tuo, M. X., Nie, J. J. 2018a, AcASn, 59, 4.

\bibitem[Zhang \& Fan(2008)] {ZF2008} Zhang, Y. W \& Fan, J. H. 2008, ChJAA, 4, 385.

\bibitem[Zhang \& Fan(2018)] {ZF2018} Zhang, L. X \& Fan, J. H. 2018, Ap\&SS, 363, 142.

\end{thebibliography}
\end{document}